\documentclass[submit]{elex2024}%%submit option for submit

\type{LETTER}

\title{An RRAM compute-in-memory architecture for high energy-efficient processing of binary matrix-vector multiplication in cryptography}

\author{Hao Yue}{1}
\author{Yihao Chen}{1}
\author{Tianhang Liang}{1}
\author{Xiangrui Li}{1}
\author{Xin Kong}{1}
\author{Zhelong Jiang}{1}
\author{Zhigang Li}{1}
\author{Gang Chen}{1}[chengang08@semi.ac.cn]
\author{Huaxiang Lu}{1}

\affiliate{1}{Institute of Semiconductors, Chinese Academy of Sciences, Beijing, 100083, China}
\vol{21}
\no{1}

\doi{10.1587/elex.XX.XXXXXXXX}
\received{2024/10/10}
\accepted{2024/10/10}
\publicized{2024/12/31}
\copyedited{2024/12/31}

\usepackage{color} 

\begin{document}

\maketitle 

\begin{abstract}
Binary matrix-vector multiplication (BMVM) is a key operation in post-quantum cryptography schemes like the Classic McEliece cryptosystem. Conventional computing architectures incur significant energy efficiency loss due to data movement of large matrices when handling such tasks. Resistive memory (RRAM) non-volatile compute-in-memory (nvCIM) is an ideal technology for high energy-efficient BMVM processing but faces challenges, including signal margin degradation in high input-parallelism arrays due to device non-idealities and high hardware overhead from current readout and XOR operations. This work presents a RRAM nvCIM architecture featuring: 1) 1T1R cells with high-resistive-state compensation modules; and 2) pulsed current-sensing parity checkers. Based on the 180nm process and test results from RRAM devices, the computing accuracy and efficiency of the architecture are verified by simulation. The proposed architecture performs high-precision current accumulation with a maximum MAC value of 10 and achieves an energy efficiency of 1.51TOPS/W, offering approximately 1.62× improvement compared to an advanced 28nm FPGA platform.
\end{abstract}



\vspace{-0.2cm}
\section{Introduction}
\vspace{-0.2cm}
Post-quantum cryptography (PQC) has attracted considerable attention in recent decades, leading to the proposal of numerous cryptographic schemes expected to resist potential threats posed by quantum computers\cite{journal paper1,journal paper2}. Binary matrix-vector multiplication (BMVM), a widely used binary logic operation in encryption, underpins many PQC candidates, including the learning parity with noise (LPN)-based cryptosystem with provable quantum security\cite{journal paper3}, and the Classic McEliece cryptosystem\cite{journal paper4}, which is a powerful contender in the PQC standardization by National Institute of Standards and Technology (NIST)\cite{website1}. However, BMVM typically involves large-scale matrix operations. Conventional von Neumann architectures, due to the separation of memory and computation, suffer from excessive power consumption and bandwidth bottleneck caused by the frequent transfer of matrix data during such operations, resulting in significant energy efficiency loss. The non-volatile compute-in-memory (nvCIM) enables direct opeations within memory arrays, effectively eliminating the energy efficiency loss caused by large-scale matrix data movement\cite{journal paper5,journal paper6,journal paper7,journal paper8,journal paper9}. This inherent advantage makes nvCIM a valid solution for energy-efficient BMVM processing. 

Resistive memory (RRAM) nvCIM has been extensively studied due to RRAM's high endurance, CMOS compatibility, high density, and low power \cite{proceeding paper1,proceeding paper2,proceeding paper3,journal paper10,proceeding paper4}. Capitalizing on their efficiency in accelerating matrix-vector multiply-accumulate (MAC) operations, numerous high-performance RRAM nvCIM designs have been proposed for neural networks computations\cite{proceeding paper5,journal paper11,proceeding paper6,journal paper12}. In certain LPN-based cryptographic protocols\cite{journal paper13} and Classic McEliece cryptosystem\cite{journal paper14}, the large matrix involved in BMVM is static, while the input vector is frequently updated. This characteristic closely aligns with the computational pattern of neural networks. Therefore, drawing inspiration from the high energy-efficient execution of neural network computations in RRAM nvCIM, designing an RRAM nvCIM architecture tailored for BMVM opeations in cryptographic applications presents a promising technological pathway. 

However, based on the aforementioned approach, achieving high precision and energy-efficient BMVM processing for cryptographic applications on RRAM nvCIM hardware still faces several challenges. The high dimensionality of the input vector denotes that the memory array necessitates parallelly inputing of multi-bit data. Due to the non-ideal resistance characteristics of the devices, the signal margins between different MAC values (MACVs) in such high input-parallel arrays will degrade. Additionally, current readout and XOR operations heavily rely on transimpedance amplifiers (TIAs) and analog-to-digital converters (ADCs), resulting in significant hardware resource consumption and quantization power overhead.

To overcome these obstacles, this paper proposes an RRAM nvCIM architecture with the following features. 

1) A 1T1R cell with a high-resistance-state (HRS) compensation module to suppress the leakage current caused by the limited HRS value. 

2) A pulsed current-sensing parity checker (PCSPC) that simultaneously satisfies the requirements of current sensing and XOR operations, eliminating the hardware overhead of TIAs and ADCs.

\vspace{-0.2cm}
\section{Computational paradigm of BMVM and challenges of its deployment on RRAM nvCIM}
\vspace{-0.2cm}
\vspace{-0.1cm}
\subsection{Computational paradigm of BMVM}
\begin{figure}[htb]
\begin{center}
\includegraphics[width=8cm]{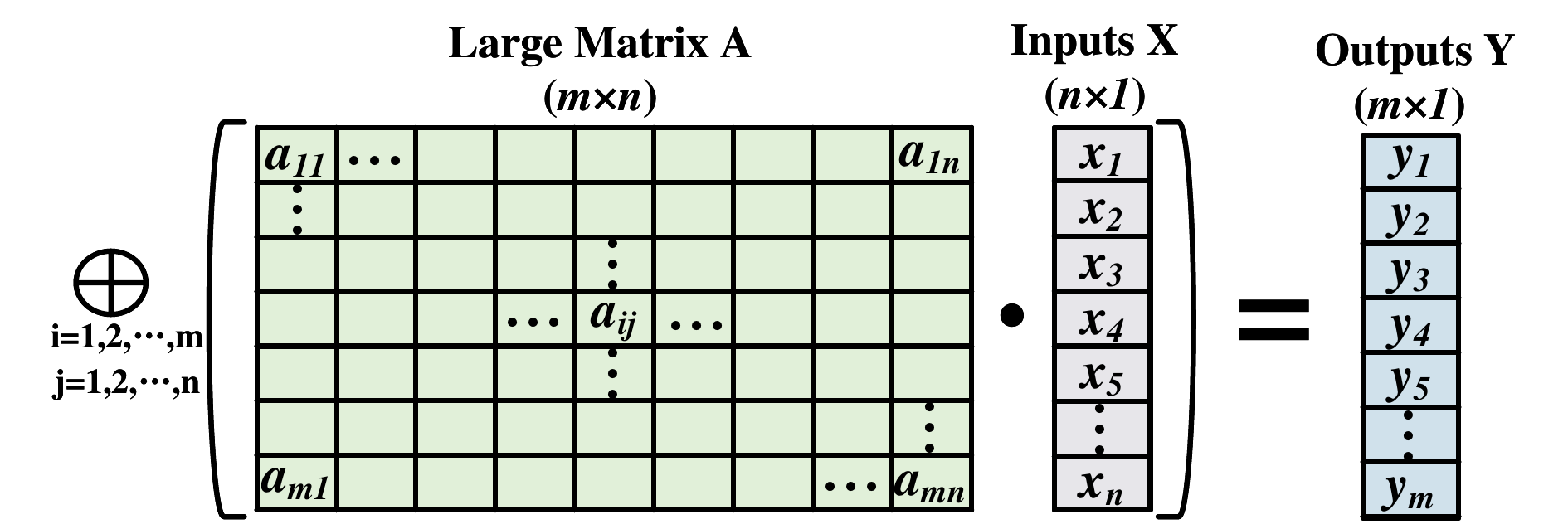}
\end{center}
\caption{The computational paradigm of BMVM.}
\label{fig:1}
\end{figure}

BMVM is characterized by simple data types and arithmetic properties, as its data is entirely defined in the binary field and requires only Boolean operations, specifically AND and XOR. BMVM primarily consists of three components: matrix \textbf{A}, vector \textbf{X}, and vector \textbf{Y}. In the cryptographic systems considered in this paper, \textbf{A} is a large static matrix, \textbf{X} represents the information to be encoded, and \textbf{Y} is the encoded output. As illustrated in Fig.~\ref{fig:1}, the computational paradigm of BMVM can be formulated as Eq.(\ref{eq1}), where \textbf{$\oplus$} denotes the XOR operation and $\cdot$ denotes the AND operation.
\begin{equation}
y_i=\underset{j=1,2 \cdots, n}{\oplus}( a_{i j} \cdot x_j), i \in\{1,2, \cdots m\}.
\label{eq1}
\end{equation}

\vspace{-0.7cm}
\vspace{-0.7cm}

\subsection{MAC signal margin degradation due to device non-idealities}
\begin{figure}[htb]
\begin{center}
\includegraphics[width=8cm]{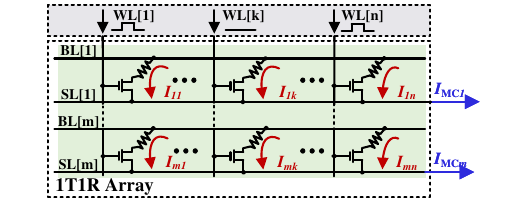}
\end{center}
\caption{Scheme of MAC Operation in a 1T1R Array  }
\label{fig:2}
\end{figure}

The 1T1R cell is widely adopted in RRAM nvCIM platforms\cite{journal paper15,proceeding paper7}, enabling  binary-input MAC operation, as shown in Fig.~\ref{fig:2}. When wordline (WL) voltage corresponds to an input of “1”, the transistor turns on, activating the cell. At this time, if RRAM in the cell stores low-resistance-state (LRS) or HRS , representing weight “1” or “0”, the cell will output a current $I_{\mathrm{LRS}}$ or $I_{\mathrm{HRS}}$, respectively. When WL voltage corresponds to an input of “0”, the cell is inactivated, thereby no current is output regardless of the RRAM state. Finally, the current from each cell is aggregated on the sourceline (SL) to obtain the accumulated current $I_{\mathrm{MC}}$. 
\vspace{-0.4cm}
\vspace{-0.4cm}

\begin{figure}[htb]
\begin{center}
\includegraphics[width=8cm]{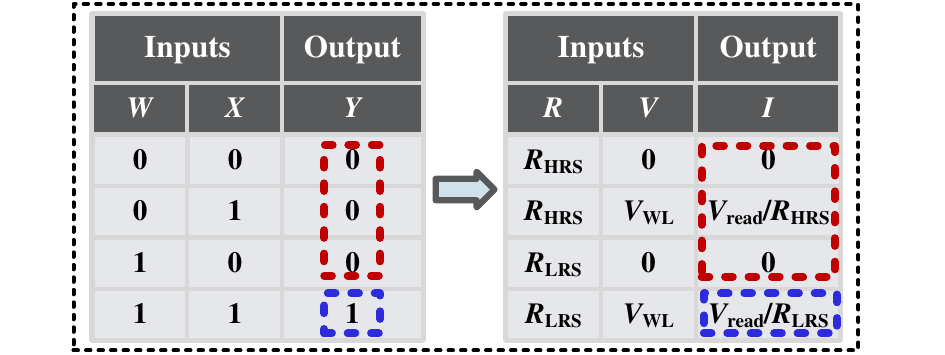}
\end{center}
\caption{Comparison of ideal and practical multiplication truth table.}
\label{fig:3}
\end{figure}

However, due to non-ideal resistance characteristics of RRAM, 1T1R arrays deviate from ideal behavior. Fig.~\ref{fig:3} shows the discrepancy between the actual and ideal cases, where  $V_{\mathrm{read}}$ represents the voltage between bitline (BL) and SL. Ideally, RRAM in the HRS is expected to reach an ultrahigh value, making the resistance ratio (R-ratio) approaches infinity and $I_{\mathrm{HRS}}$ is nearly zero. Nevertheless, the HRS resistance varies significantly during programming, making it difficult to reliably achieve such ultrahigh values. The limited HRS resistance results in small R-ratio and a non-negligible leakage current $V_{\mathrm{read}}/R_{\mathrm{HRS}}$ when the cell is activated. As the number of cells involved in the MAC operation increases with the input vector dimension, leakage currents and small R-ratio lead to the degradation of $I_{\mathrm{MC}}$ signal margin. $I_{\mathrm{MC}}$ corresponding to different MACVs will be wide-spread and narrow-spaced, even exhibiting significant overlap, so that different MACVs can not be distinguished\cite{journal paper16}. In addition, the LRS of RRAM also exhibits small programming variability, resulting in $I_{\mathrm{LRS}}=V_{\mathrm{read}}/R_{\mathrm{LRS}}$ not being a perfectly constant value. This further exacerbates the degradation of signal margin, leading to an increased computation bit error rate (BER).

\setlength{\textfloatsep}{8pt} 

\subsection{High hardware overhead from current readout and XOR operations}
The addition in BMVM differs from conventional accumulation, as it follows modulo-2 addition (XOR logic operation). Thus, $I_{\mathrm{MC}}$ must be converted into the result of XOR operation, i.e., the implementation of XOR activation function. The combination of TIA and ADC is commonly used current-sensing schemes in RRAM nvCIM platforms\cite{proceeding paper8,proceeding paper9}. TIA receives $I_{\mathrm{MC}}$ and proportionally convert it into a voltage, which is then sampled and quantized by ADC into a multi-bit digital code. This readout scheme can also be leveraged for the implementation of XOR activation function. Based on the working principle of ADC, the least significant bit (LSB) of the output digital code essentially represents the result of  XOR logic operation. 

Although the above scheme provides an effective solution for implementing XOR activation function, it heavily relies on TIA and ADC, causing significant hardware overhead in the large-scale array. Moreover, since ADC typically outputs the LSB in the final stage of the conversion cycle, many unnecessary intermediate bits are generated, leading to substantial quantization power consumption. 

\vspace{-0.2cm}
\section{Proposed RRAM nvCIM design}
\vspace{-0.2cm}
\vspace{-0.1cm}
\subsection{Overall architecture of proposed RRAM nvCIM}
\begin{figure}[htb]
\begin{center}
\includegraphics[width=8cm]{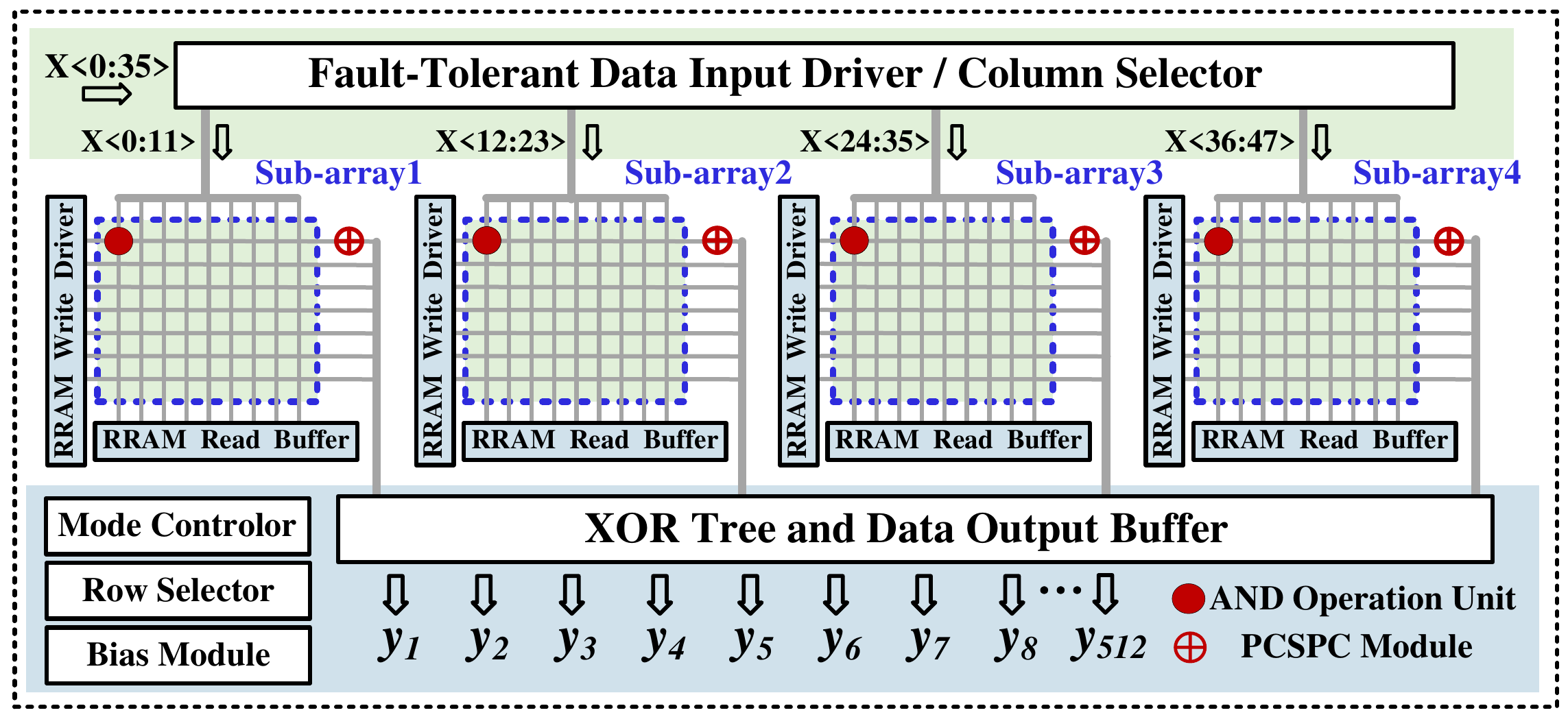}
\end{center}
\caption{Overall architecture of proposed RRAM nvCIM design.}
\label{fig:4}
\end{figure}

This section details the proposed RRAM nvCIM architecture, as shown in Fig.~\ref{fig:4}. It primarily comprises a fault-tolerant data input driver, a column and row selector, a mode controller, a bias module, an XOR tree, an output buffer, and four RRAM sub-arrays. The fault-tolerant data input driver and output buffer are used to perform data input and output. The column and row selector are used to control RRAM write driver and RRAM read buffer to facilitate weight writing and reading, where the column selector is implemented by reusing the fault-tolerant data input driver to save hardware resource. The mode controller determines whether the circuit operates in memory or CIM mode. The XOR tree and RRAM sub-arrays form the computational core, responsible for performing the AND and XOR operations involved in BMVM. Each sub-array contains multiple AND operation units, and each row within a sub-array is equipped with a PCSPC module. The bias module provides the bias voltage to AND operation units within the array.

\setlength{\textfloatsep}{6pt} 

\begin{figure}[htb]
\begin{center}
\includegraphics[width=8cm]{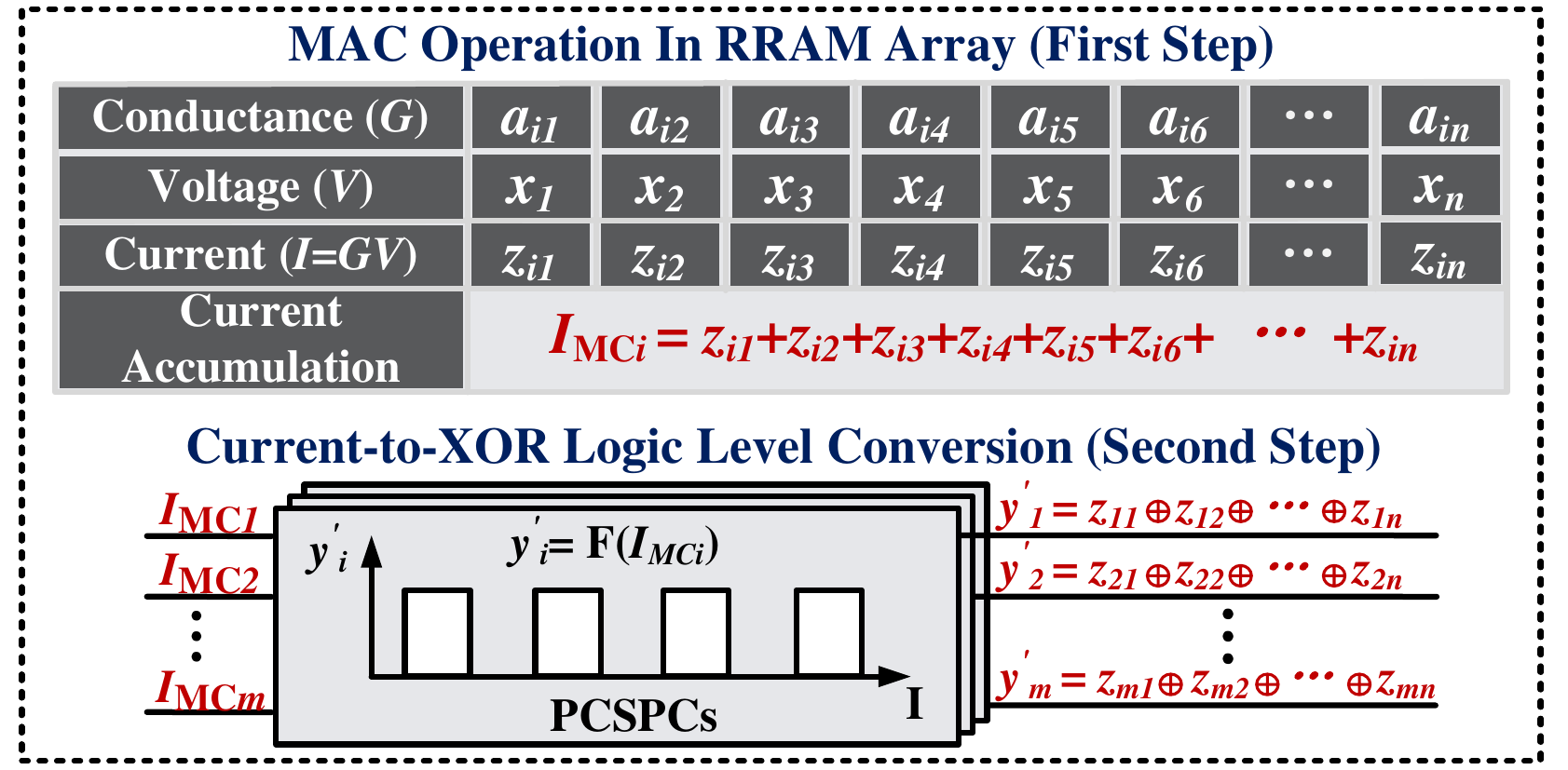}
\end{center}
\caption{Dataflow for BMVM computation in the proposed array.}
\label{fig:5}
\end{figure}

When deploying BMVM operations on the proposed architecture, input voltage and output voltage pulses represent the data of \textbf{X} and \textbf{Y}, respectively. A high voltage pulse denotes logic “1”, whereas the absence of a pulse indicates logic “0”. The elements of large matrix \textbf{A} are mapped to the conductance values (i.e., the reciprocal of resistance values) of RRAM. When $a_{ij}$ = 1, RRAM is programmed to high-conductance-state; when $a_{ij}$ = 0, it is programmed to low-conductance-state. Taking the deployment of a M×N BMVM task on a single RRAM sub-array as an example, Fig.~\ref{fig:5} illustrates the dataflow of the RRAM sub-array during BMVM execution. The first step is the MAC operation. Each AND operation unit in the RRAM sub-array performs the logical AND between $a_{ij}$ and $x_{j}$, and outputs current $z_{ij}$. When both $a_{ij}$ and $x_{j}$ equal 1, the output current $z_{ij}$ is approximately 4µA; otherwise, it is approximately 0µA. After that,  the output currents from different units in a row will be accumulated together. The second step is that the PCSPC module equipped in each row receives $I_{\mathrm{MC}}$ and transforms it into a logic level $y_{i}^{'}$ corresponding to XOR result. As all four sub-arrays operate simultaneously, a simple XOR tree module can be employed to merge the outputs $y_{i}^{'}$ from the four sub-arrays, yielding the final XOR result $y_{i}$.
\vspace{-0.4cm}
\vspace{-0.4cm}

Each RRAM sub-array contains 512×12 AND units, with 9 columns for computation and 3 columns as redundancy. Two redundant bits are inactivated, and the fault-tolerant data input driver can dynamically select which columns are designated as inactive redundancy, thereby enhancing the array's tolerance to RRAM device yield variations. The last redundant bit is configured to continuously output a current of approximately 4µA, which accelerates the circuit establishment process. As a result, each sub-array is capable of handling a 512×9 computation task, and the four sub-arrays collectively support a 512×36 computation task. Thanks to the scalability of the sub-array structure, larger-scale tasks can be supported simply by increasing the number of sub-arrays or extending rows of per sub-array, along with implementing a deeper XOR tree. Meanwhile, each sub-array operates in a highly parallel manner, ensuring that the overall operating frequency of the circuit remains largely unaffected by the increase of task size. This strategy of dividing an entire array into multiple sub-arrays effectively limits the maximum MACVs to 10, even when handling computation tasks of larger-scale, while maintaining high throughput. This effectively mitigates the degradation of signal margin when processing input vectors with high dimensionality. 
\vspace{-0.2cm}
\subsection{Structure of AND operation unit}
\begin{figure}[htb]
\begin{center}
\includegraphics[width=8cm]{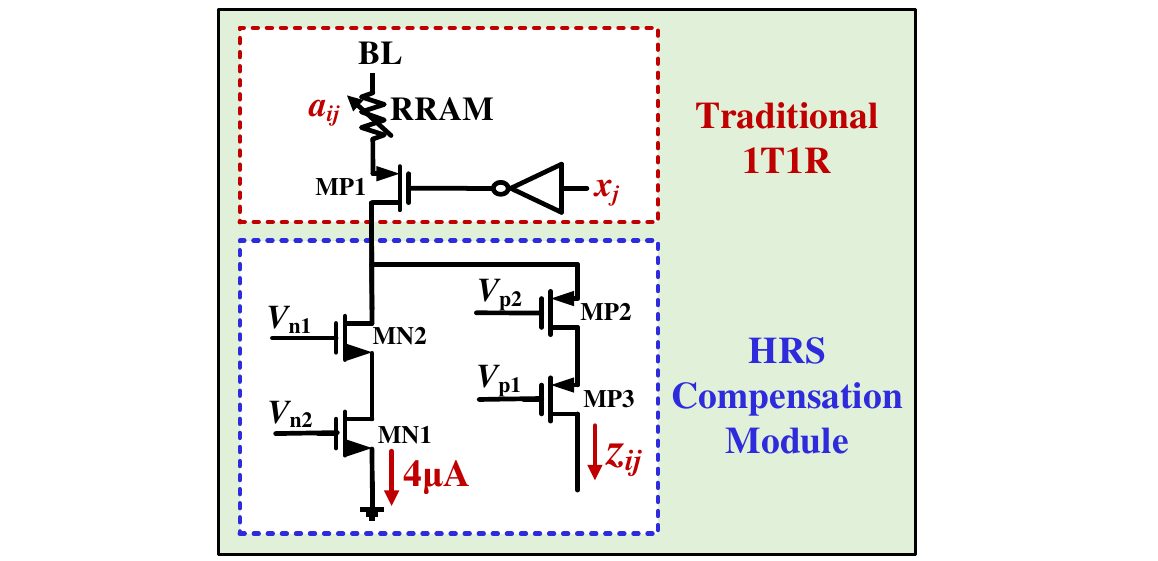}
\end{center}
\caption{The Structure of 1T1R cell with HRS compensation.}
\label{fig:6}
\end{figure}

To mitigate the issue of leakage current, the proposed AND operation unit enhances the traditional 1T1R structure by incorporating an HRS compensation module with two branches formed by four transistors, as shown in Fig.~\ref{fig:6}. When the AND operation unit is activated, the PMOS transistor MP1 is turned on. At this time, under the control of the bias module, the left branch (comprising two NMOS transistors MN1 and MN2) is designed to provide a constant current of 4µA, and the right branch (comprising two PMOS transistors MP2 and MP3) is designed to output a 4µA current representing $z_{ij}$ = 1 when the RRAM in LRS. Since the 4µA current from the left branch constantly flows through the RRAM regardless of its state, a higher voltage drop is consumed across RRAM in HRS compared to that in LRS. Therefore, when the RRAM is programmed to the HRS, the gate-source voltage of MP2 significantly decreases, driving it into the subthreshold region and thereby restricting the output current of right branch to approximately 0µA. With the inclusion of HRS compensation module, the operation unit no longer relies solely on the high resistance value of RRAM in HRS to suppress leakage current. Instead, by actively forcing the output transistor into subthreshold region, the design effectively improves the equivalent R-ratio of RRAM and leakage current suppression capability of the unit. It is noteworthy that a larger constant current in the left branch leads to a lower gate-source voltage of MP2 when the RRAM is in HRS, thereby enhancing the leakage current suppression capability of the cell. However, a larger current typically results in higher power consumption. By setting the constant current in the left branch to 4µA, the circuit achieves a favorable trade-off between strong leakage suppression and low power consumption.
\vspace{-0.4cm}
\vspace{-0.8cm}
\subsection{Implementation of PCSPC module}
\begin{figure}[htb]
\begin{center}
\includegraphics[width=8cm]{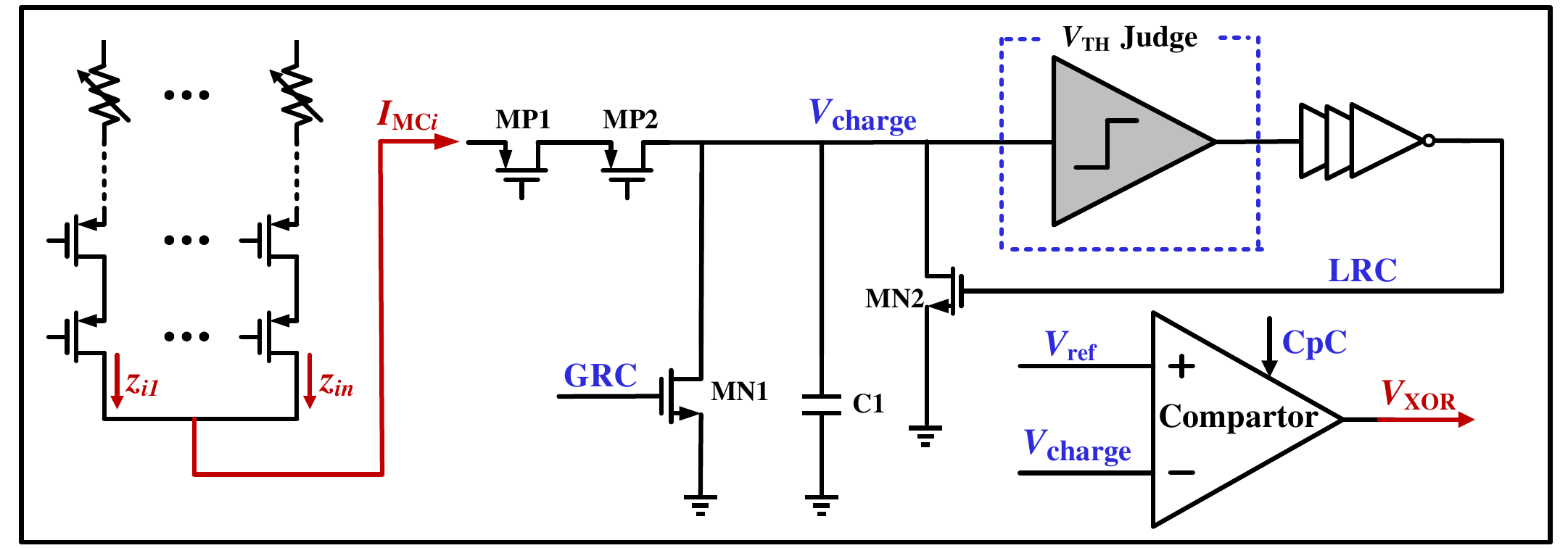}
\end{center}
\caption{Overall Structure of PCSPC module.}
\label{fig:7}
\end{figure}

Fig.~\ref{fig:7} shows the implementation scheme of the PCSPC module. The accumulated current $I_{\mathrm{MC}}$ flows through PMOS transistors MP1 and MP2 to charge capacitor C1, thereby increasing voltage $V_{\mathrm{charge}}$, with MP1 and MP2 mitigating the impact of capacitor charging on the accuracy of $I_{\mathrm{MC}}$. The $V_{\mathrm{TH}}$ judge module can generate local reset clock (LRC) by detecting whether the value of $V_{\mathrm{charge}}$ exceeds the threshold $V_{\mathrm{TH}}$, without requiring an external $V_{\mathrm{TH}}$ input. The NMOS transistors MN1 and MN2 form the global and local reset module, respectively, to discharge the C1. The comparator performs a comparison between the $V_{\mathrm{charge}}$ and the reference voltage $V_{\mathrm{ref}}$ when the comparator clock (CpC) is active.

\begin{figure}[htb]
\begin{center}
\includegraphics[width=8cm]{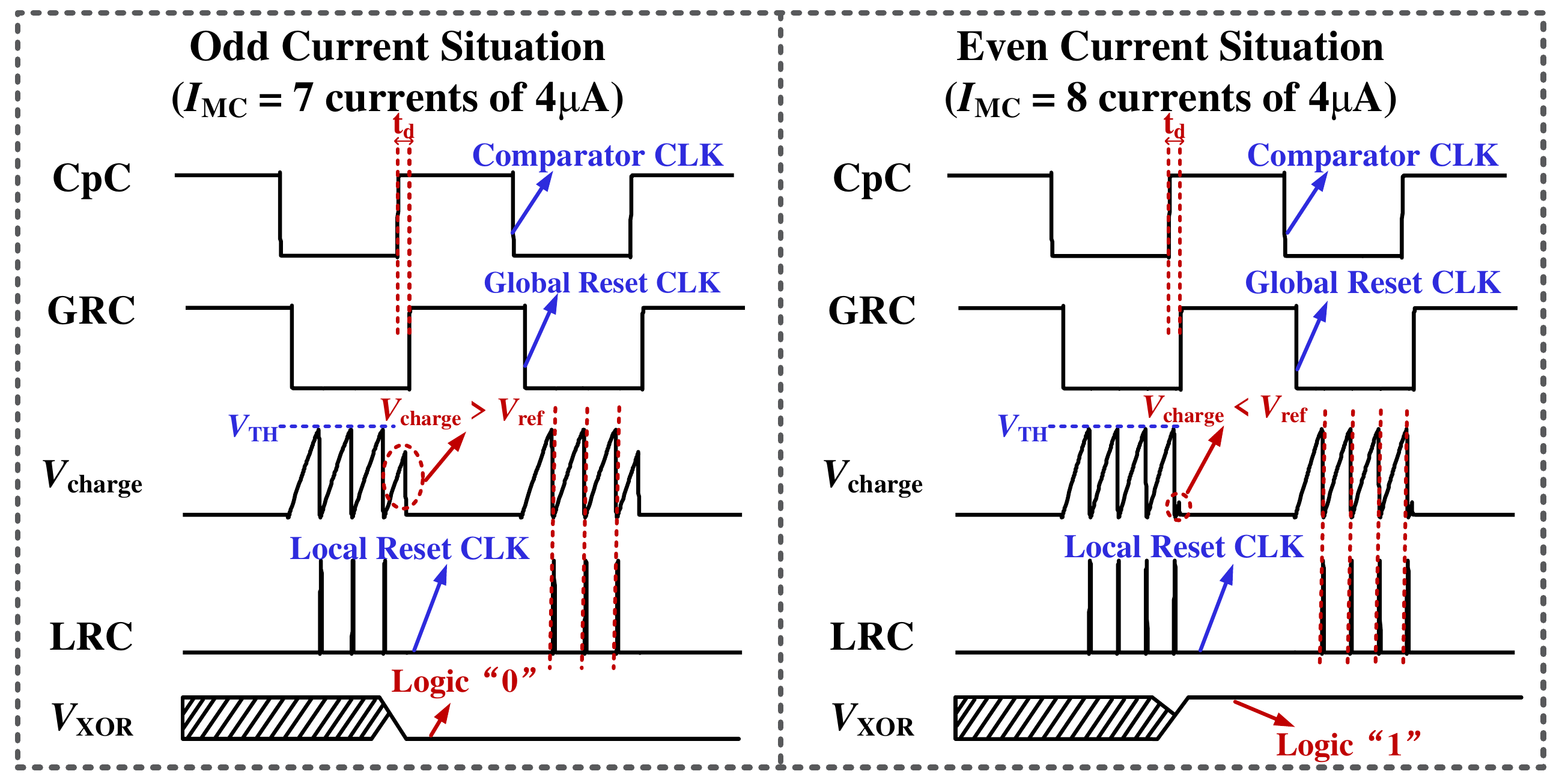}
\end{center}
\caption{Operating waveforms of the PCSPC module.}
\label{fig:8}
\end{figure}

\setlength{\textfloatsep}{6pt}
The PCSPC module is designed based on an intuitive observation that the XOR value depends on the parity of Hamming weight of $I_{\mathrm{MC}}$. If the Hamming weight is odd, the XOR value is 1, and if that is even, the XOR value is 0. Fig.~\ref{fig:8} illustrates the operating waveforms of PCSPC module when Hamming weight of $I_{\mathrm{MC}}$ is 7 (odd) and 8 (even), respectively. During the global reset clock (GRC) is low, the accumulated current charges C1, and under the coordination of $V_{\mathrm{TH}}$ judge module and MN2, $V_{\mathrm{charge}}$ exhibits a certain number of ramp pulses. By properly designing the period of GRC and the capacitance value, the number of generated ramp pulses equals half Hamming weight of $I_{\mathrm{MC}}$. Therefore, when the Hamming weight is odd, the value of  $V_{\mathrm{charge}}$ just before GRC transitions high is approximately half of $V_{\mathrm{TH}}$; otherwise, it is approximately zero.The comparator is triggered when CpC goes high, which is designed to precede GRC by a short time interval $t_{\mathrm{d}}$. As a result, the comparator compares $V_{\mathrm{charge}}$ just before GRC transitions high with $V_{\mathrm{ref}}$ being set approximately to the midpoint between zero and half of $V_{\mathrm{TH}}$. If $V_{\mathrm{charge}}$ at this time exceeds $V_{\mathrm{ref}}$, it indicates that $I_{\mathrm{MC}}$ has accumulated an odd number of 4µA currents, thereby the XOR value is determined to be 1; otherwise, the XOR value is determined to be 0. Although the variability of RRAM resistance may cause some deviation in the $V_{\mathrm{charge}}$  just before GRC transitions high, the $V_{\mathrm{ref}}$ remains sufficient distance from both half of $V_{\mathrm{TH}}$ and zero. This provides the circuit with a high tolerance to RRAM resistance variations, thereby ensuring the accuracy of the XOR value. Additionally, it is worth noting that due to the presence of a redundant bit with constant 4µA output current, the XOR value needs to be inverted to obtain the final correct output $V_{\mathrm{XOR}}$.
\vspace{-0.4cm}
\vspace{-0.4cm}
\vspace{-0.4cm}
\section{Experiment}
\vspace{-0.2cm}
\vspace{-0.1cm}
\subsection{Testing and modeling of RRAM devices}
\begin{figure}[h]
\begin{center}
\includegraphics[width=8cm]{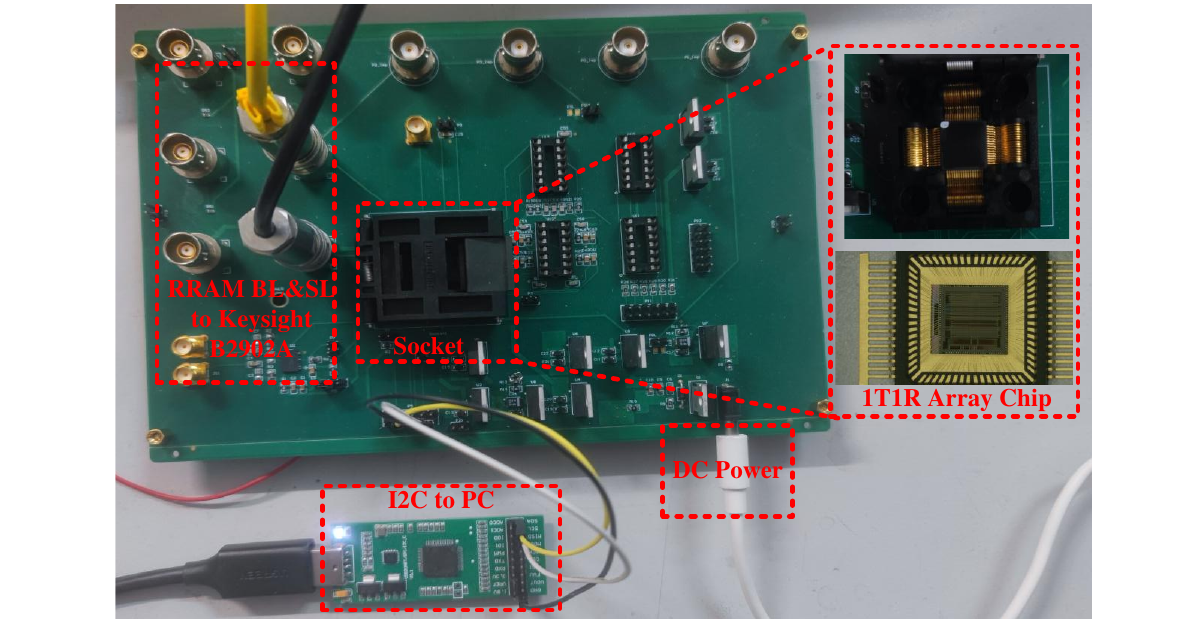}
\end{center}
\caption{Micrograph and test board of the 1T1R array chip.}
\label{fig:9}
\end{figure}

\begin{figure}[h]
\begin{center}
\includegraphics[width=8cm]{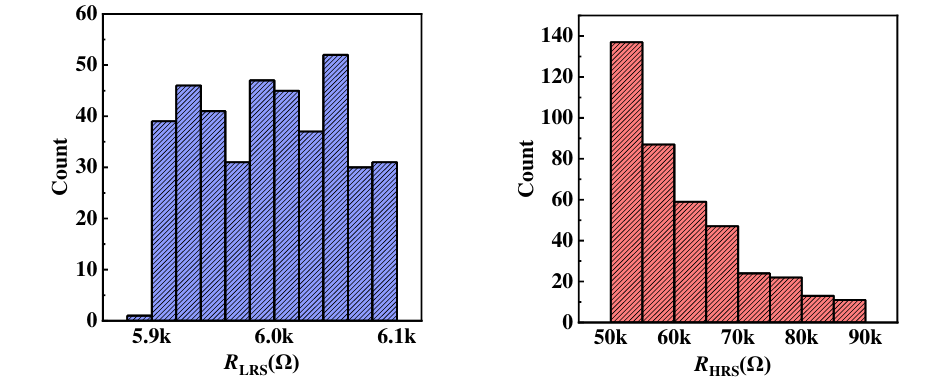}
\end{center}
\caption{High and low resistance distribution test results of the RRAM.}
\label{fig:10}
\end{figure}

\begin{figure}[h]
\begin{center}
\includegraphics[width=8cm]{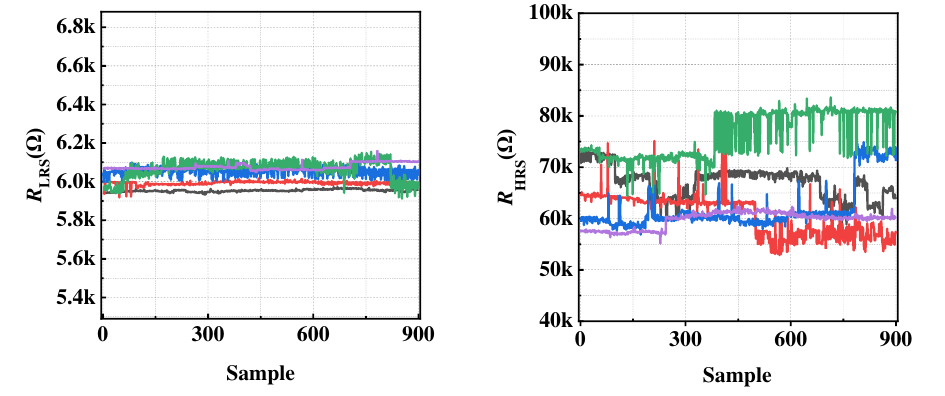}
\end{center}
\caption{Retention characteristic test results of RRAM resistance states.}
\label{fig:11}
\end{figure}

To evaluate the resistance non-idealities of RRAM, a 1T1R array chip was tested, with the chip micrograph and test board shown in Fig.~\ref{fig:9}. In this experiment, multiple RRAM underwent repeated programming between HRS and LRS, and the resulting resistance values were statistically analyzed. As shown in Fig.~\ref{fig:10}, the LRS can remained stable around 6k with less than 2\% variation, while the HRS exceeds 50k with a significantly larger variation range of about 40k. In addition, to verify the feasibility of employing RRAM in nvCIM, the resistance retention characteristics were evaluated. Specifically, multiple read operations were performed on both the HRS and LRS of several programmed devices and the results are shown in Fig.~\ref{fig:11}. It was observed that both resistance states remained stable during a long period of read operations. Although the HRS exhibited relatively large fluctuations during read operations, its resistance consistently stayed above a high reference level, which is sufficient to meet the requirements of CIM.

\vspace{-0.4cm}
\vspace{-0.9cm}
\vspace{-0.9cm}
\vspace{-0.4cm}

\begin{figure}[!h]
\begin{center}
\includegraphics[width=8cm]{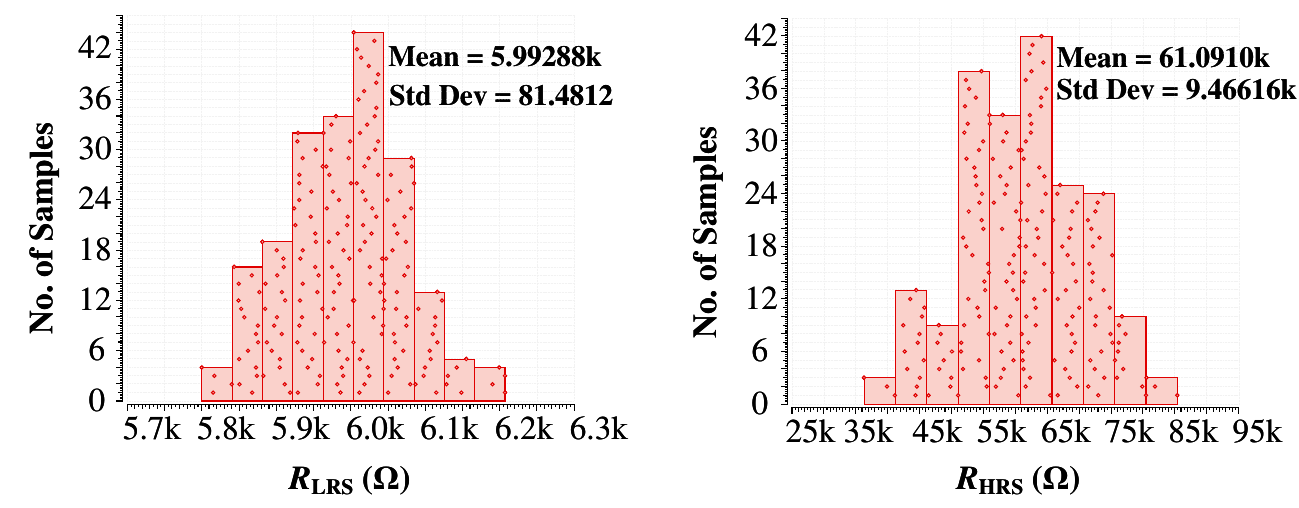}
\end{center}
\caption{Monte Carlo simulation results of the established model.}
\label{fig:12}
\end{figure}

Based on the results shown in Fig.~\ref{fig:10}, a Gaussian distribution model of RRAM resistance was established and integrated into the EDA tool for subsequent performance evaluation simulations. Although the actual resistance distribution of RRAM does not strictly follow a centralized Gaussian trend, the key factors affecting circuit performance are the low values of HRS and the fluctuation range of LRS, rather than the overall resistance distribution shape. The Gaussian model, characterized by its variance parameters, effectively captures the spread of both resistance states, providing a quantitative basis for analyzing the impact of resistance non-idealities on circuit performance. The Monte Carlo simulation results of the model are shown in Fig.~\ref{fig:12}, where LRS fluctuation is slightly larger than the measured results, and the minimum value of HRS is over 10k lower than that in the measurement. Compared with the test data, the modeled resistance exhibits more severe non-idealities, thereby serving as a conservative and effective mean to validate whether the circuit meets computational requirement. 
\vspace{-0.8cm}
\vspace{-0.8cm}

\subsection{Computation accuracy evaluation  }
\begin{figure}[htb]
\begin{center}
\includegraphics[width=8cm]{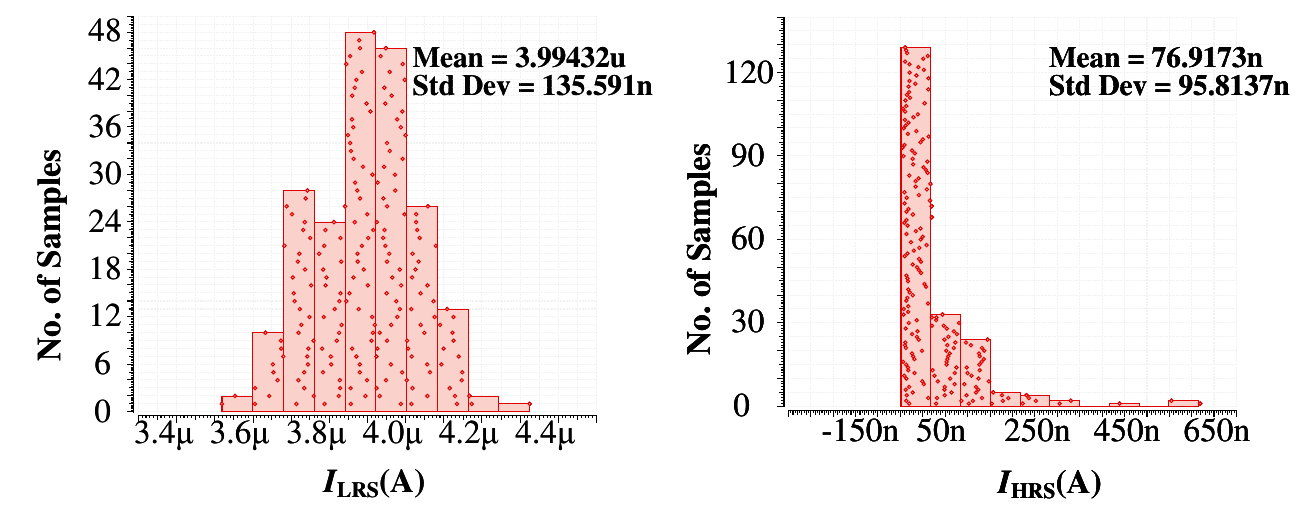}
\end{center}
\caption{Monte Carlo simulation results of $I_{\mathrm{LRS}}$ and $I_{\mathrm{HRS}}$.}
\label{fig:13}
\end{figure}

Fig.~\ref{fig:13} shows the Monte Carlo simulation results of $I_{\mathrm{LRS}}$ and $I_{\mathrm{HRS}}$ for the AND operation unit based on the established model.  It can be observed that proposed unit effectively suppresses leakage current and achieves an equivalent average R-ratio of approximately 51.9, representing approximately 5× improvement compared to the 1T1R.
\vspace{-0.4cm}
\vspace{-0.4cm}

\begin{figure}[htb]
\begin{center}
\includegraphics[width=8cm]{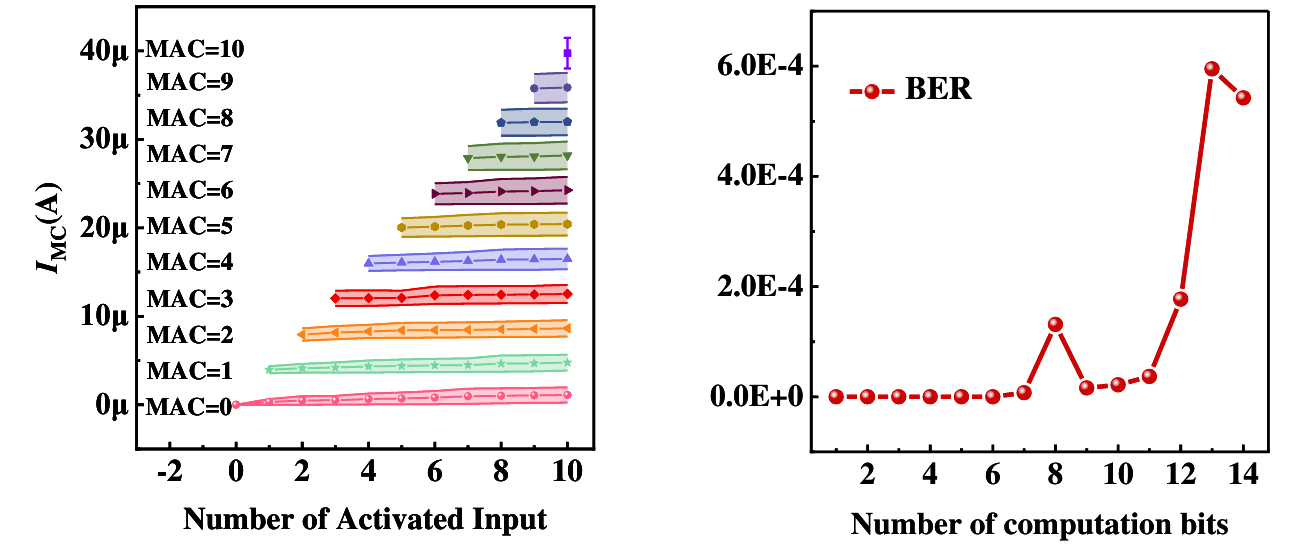}
\end{center}
\caption{Evaluation of MAC operation accuracy and BER.}
\label{fig:14}
\end{figure}

Considering the redundant bit, the maximum MACV in the proposed architecture is 10. Fig.~\ref{fig:14} (left) illustrates the current fluctuation range for each MACV via Monte Carlo simulation. Taking MACV equal to 6 as an example, 6 out of the 10 AND operation units are activated and output a current $I_{\mathrm{LRS}}$, while the remaining 4 cells are either in an inactive state with no output current or in an activated state but output a leakage current $I_{\mathrm{HRS}}$. As the number of activated input increases, the number of cells among these 4 that output leakage current also increases. Therefore, when MACV equals 6, there are five possible scenarios for the $I_{\mathrm{MC}}$, as shown in Fig.~\ref{fig:14} (left). By performing Monte Carlo simulations for each scenario where MACV equals 6, the fluctuation range of the corresponding $I_{\mathrm{MC}}$ can be obtained. It can be observed from the figure that even when extremely rare current values are taken into account, all MAC signals maintain a positive margin, with no overlap between adjacent current levels.

\vspace{-0.4cm}
\vspace{-0.4cm}

By performing Monte Carlo simulations on the output results of the PCSPC modules within sub-arrays configured with different numbers of computation bits, the BER of sub-array with different numbers of computation bits can be obtained by calculating the ratio of erroneous output points to the total number of points, as shown in  Fig.~\ref{fig:14} (right). For the proposed architecture, each sub-array operates with 9 computation bits, yielding a BER of approximately 1.6E-5. Taking the authentication protocol based on biometric LPN commitments proposed in \cite{journal paper13} as a use case, the impact of BER of 1.6E-5 on the protocol’s performance, including the false acceptance rate (FAR) and false rejection rate (FRR), was evaluated based on the FV3\_Test subset of the THU-FVFDT3 finger vein dataset from Tsinghua University\cite{website2}. Experimental results show that, compared to the protocol without induced BER, the protocol with a BER of 1.6E-5 exhibited no increase in FAR, which remained at 0\%. However, the FRR increased by approximately 10\%. Fortunately, the elevated FRR still remained at a low level, and the protocol supports repeated authentication attempts, which further enhances its tolerance to increased FRR. Therefore, it can be concluded that a BER of 1.6E-5 is acceptable for practical cryptographic systems such as the one in \cite{journal paper13}.

\vspace{-0.2cm}
\subsection{Computation efficiency evaluation}
Based on the physical architecture of the proposed design, the PCSPC module is required to distinguish 10 discrete current levels, which is equivalent to achieving a resolution of approximately 3.3-bit in conventional ADC-based  scheme. The operating frequency of the PCSPC module is primarily determined by integration capacitor. From the perspective of linearity and through simulation, the optimal value of the capacitor can be obtained, which defines the PCSPC module’s operating frequency at 40MHZ. In addition, by simulating the product of voltage and average current, the power consumption of the PCSPC module under different input current levels can be evaluated. Taking into account the probability distribution of each current level, the overall power consumption of the PCSPC module is calculated to be 0.097mW. Table~\ref{tab:1} presents a performance comparison between the PCSPC module and several 4-bit ADCs. Although the PCSPC module operates at a relatively low frequency, it demonstrates a significant advantage in terms of power. 

\vspace{-0.4cm}
\begin{table}[ht]
\small
\caption{Performance comparison of the PCSPC Module } \label{tab:1}
\centering
\tabcolsep.9\tabcolsep
\begin{tabular}{ccccc}
\hline
Work& Paper\cite{journal paper17}&  Paper\cite{proceeding paper10}&Paper\cite{journal paper18}&{\bf This Work}\\
\hline
Process(nm)& 180&  130& 65&{\bf 180}\\
\hline
Resolution(bits)& 4&  4& 4&{\bf3.3}\\
\hline
Frequency(MHZ)& 1600&  62& 500&{\bf 40}\\
\hline
Power(mW)& 15.5&  1& 78&{\bf 0.097}\\
\hline
\end{tabular}
\end{table}
\vspace{-0.4cm}

To further evaluate the computation performance of the proposed RRAM nvCIM architecture, the same computing task with a scale of 512×36 was deployed on both an FPGA platform and the nvCIM platform, followed by simulation-based experiments. The FPGA platform adopts the Xilinx KC705 development board, which integrates XC7K325T-2FFG900C chip fabricated with 28nm process, and experiment on FPGA platform is conducted using a vector-based evaluation methodology. Table~\ref{tab:2} presents the performance comparison between two hardware platforms. Given the trade-off between power consumption and throughput, energy efficiency was selected as the primary metric for computational performance. The results show that, despite being implemented using 180nm process, the proposed RRAM nvCIM platform still achieves approximately 1.62× improvement in energy efficiency compared to the advanced FPGA platform based on 28nm technology. 

\vspace{-0.4cm}
\begin{table}[ht]
\small
\caption{Performance comparison of proposed RRAM nvCIM platform } \label{tab:2}
\centering
\tabcolsep.9\tabcolsep
\begin{tabular}{ccc}
\hline
Platform& FPGA&  {\bf This Work}\\
\hline
Process(nm)& 28&  {\bf 180}\\
\hline
BER& $\sim$ 0&  {\bf 1.6E-5}\\
\hline
Power(W)& 1.975&  {\bf 0.487}\\
\hline
throughput(Gpbs)& 51.20& {\bf 20.48}\\
\hline
Energy Efficiency(TOPS/W)& 0.93& {\bf 1.51}\\
\hline
\end{tabular}
\end{table}
\vspace{-0.8cm}

\section{Conclusion}
\vspace{-0.3cm}

This paper proposes an RRAM nvCIM architecture for high energy-efficient BMVM processing. To mitigate MAC signal margin degradation in high input-parallel arrays, an HRS compensation module is designed to increase the equivalent R-ratio and suppresses leakage current. Furthermore, a PCSPC module is developed to support simultaneously current sensing and XOR computation, significantly reducing hardware and power overhead. Experimental results validate the superiority of the proposed RRAM nvCIM platform.

\vspace{-0.2cm}
\section*{Acknowledgments}
\vspace{-0.3cm}

This work was supported in part by the CAS Strategic Leading Science and Technology Project XDB44000000 and in part by the National Natural Science Foundation of China 92364202.


\vspace{-0.2cm}
\begin{thebibliography}{99}% 9 or 99
\vspace{-0.3cm}
\bibitem{journal paper1} 
D. J. Bernstein and T. Lange: ``Post-quantum cryptography, " Nature {\bf 549} (2017) 7671 (DOI:10.1038/nature23461).

\bibitem{journal paper2} 
D. Joseph,  {\it et al.}: ``Transitioning organizations to post-quantum cryptography,'' Nature, {\bf 605} (2022) 7909 (DOI: 10.1038/s41586-022-04623-2).

\bibitem{journal paper3} 
A. Esser, {\it et al.}: ``LPN decoded,'' Annual International Cryptology Conference (2017) (DOI: 10.1007/978-3-319-63715-0-17).

\bibitem{journal paper4} 
R. J. McEliece: ``A public-key cryptosystem based on algebraic coding theory," Deep Space Network Progress Report, (1978). 

\bibitem{website1}
NIST (2022) https://csrc.nist.gov/news/2022/pqc-candidates-to-be-standardized-and-round-4.

\bibitem{journal paper5}
M. Lanza, {\it et al.}: ``Memristive technologies for data storage, computation, encryption, and radio-frequency communication,''  Science {\bf 376} (2022) 6597 (DOI: 10.1126/science.abj9979).

\bibitem{journal paper6}
W. H. Chen, {\it et al.}: ``CMOS-integrated memristive non-volatile computing-in-memory for AI edge processors,'' Nature Electronics {\bf 2} (2019) 9 (DOI: 10.1038/s41928-019-0288-0).

\bibitem{journal paper7}
H. S. P. Wong and S. Salahuddin S: ``Memory leads the way to better computing,'' Nature nanotechnology {\bf 10} (2015) 3 (DOI: 10.1038/nnano.2015.29).

\bibitem{journal paper8}
C. X. Xue, {\it et al.}: ``Embedded 1-Mb ReRAM-based computing-in-memory macro with multibit input and weight for CNN-based AI edge processors,'' IEEE Journal of Solid-State Circuits {\bf 55} (2019) 1 (DOI: 10.1109/JSSC.2019.2951363). 

\bibitem{journal paper9}
J. M. Hung, {\it et al.}: ``A four-megabit compute-in-memory macro with eight-bit precision based on CMOS and resistive random-access memory for AI edge devices," Nature Electronics {\bf 4} (2021) 12 (DOI: 10.1038/s41928-021-00676-9).

\bibitem{proceeding paper1}
G. Sassine, {\it et al.}: ``Sub-pJ consumption and short latency time in RRAM arrays for high endurance applications," 2018 IEEE International Reliability Physics Symposium (IRPS). IEEE (2018) (DOI: 10.1109/IRPS.2018.8353675).

\bibitem{proceeding paper2}
R. Fackenthal, {\it et al.}: ``19.7 A 16Gb ReRAM with 200MB/s write and 1GB/s read in 27nm technology," 2014 IEEE International Solid-State Circuits Conference Digest of Technical Papers (ISSCC). IEEE (2014) (DOI: 10.1109/ISSCC.2014.6757460).

\bibitem{proceeding paper3}
L. Wang L, {\it et al.}: ``A 14nm 100Kb 2T1R Transpose RRAM with> 150X resistance ratio enhancement and 27.95\% reduction on energy-latency product using low-power near threshold read operation and fast data-line current stabling scheme," 2021 Symposium on VLSI Technology. IEEE (2021).

\bibitem{journal paper10}
H. Y. Lee, {\it et al.}: ``Low-power and nanosecond switching in robust hafnium oxide resistive memory with a thin Ti cap," \textit{IEEE Electron Device Letters} {\bf 31} (2009) 1 (DOI:10.1109/LED.2009.2034670).  

\bibitem{proceeding paper4}
S. D. Spetalnick, {\it et al.}: ``A 40nm 64kb 26.56 tops/w 2.37 mb/mm 2 rram binary/compute-in-memory macro with 4.23 x improvement in density and> 75\% use of sensing dynamic range," 2022 IEEE International Solid-State Circuits Conference (ISSCC). IEEE (2022) (DOI: 10.1109/ISSCC42614.2022.9731725).   

\bibitem{proceeding paper5}
C. X. Xue, {\it et al.}: ``24.1 A 1Mb multibit ReRAM computing-in-memory macro with 14.6 ns parallel MAC computing time for CNN based AI edge processors," 2019 IEEE International Solid-State Circuits Conference-(ISSCC). IEEE(2019) (DOI: 10.1109/ISSCC.2019.8662395).   

\bibitem{journal paper11}
S. Jeong, {\it et al.}: ``Variation-tolerant and low R-ratio compute-in-memory ReRAM macro with capacitive ternary MAC operation," \textit{IEEE Transactions on Circuits and Systems I: Regular Papers} {\bf 69} (2022) (DOI: 10.1109/TCSI.2022.3165352).

\bibitem{proceeding paper6}
C. X. Xue, {\it et al.}: ``15.4 A 22nm 2Mb ReRAM compute-in-memory macro with 121-28TOPS/W for multibit MAC computing for tiny AI edge devices," 2020 IEEE International Solid-State Circuits Conference-(ISSCC). IEEE (2020) (DOI: 10.1109/ISSCC19947.2020.9063078).


\bibitem{journal paper12}
S. Zhang, {\it et al.}: ``A robust 8-bit non-volatile computing-in-memory core for low-power parallel MAC operations," \textit{IEEE Transactions on Circuits and Systems I: Regular Papers} {\bf 67} (2020) 6 (DOI: 10.1109/TCSI.2020.2971642).

\bibitem{journal paper13}
R. Arjona and I. Baturone: ``A post-quantum biometric template protection scheme based on learning parity with noise (LPN) commitments," \textit{IEEE Access} {\bf 8} (2020) (DOI: 10.1109/ACCESS.2020.3028703).

\bibitem{journal paper14}
R. Arjona, {\it et al.}: ``Post-quantum biometric authentication based on homomorphic encryption and classic McEliece," \textit{Applied Sciences} {\bf 13} (2023) 2 (DOI: 10.3390/app13020757).

\bibitem{journal paper15}
J. M. Hung, {\it et al.}: ``A four-megabit compute-in-memory macro with eight-bit precision based on CMOS and resistive random-access memory for AI edge devices," Nature Electronics {\bf 4} (2021) 12 (DOI: 10.1038/s41928-021-00676-9).

\bibitem{proceeding paper7}
W. H. Chen, {\it et al.}: ``A 65nm 1Mb nonvolatile computing-in-memory ReRAM macro with sub-16ns multiply-and-accumulate for binary DNN AI edge processors," 2018 IEEE International Solid-State Circuits Conference-(ISSCC). IEEE (2018) (DOI: 10.1109/ISSCC.2018.8310400)

\bibitem{journal paper16}
L. Wang, {\it et al.}: ``Efficient and robust nonvolatile computing-in-memory based on voltage division in 2T2R RRAM with input-dependent sensing control," IEEE Transactions on Circuits and Systems II: Express Briefs {\bf 68} (2021) 5 (DOI: 10.1109/TCSII.2021.3067385). 

\bibitem{proceeding paper8}
J. M. Correll, {\it et al.}: ``An 8-bit 20.7 TOPS/W multi-level cell ReRAM-based compute engine," 2022 IEEE Symposium on VLSI Technology and Circuits (VLSI Technology and Circuits). IEEE (2022) (DOI: 10.1109/VLSITechnologyandCir46769.2022.9830490) 

\bibitem{proceeding paper9}
G. Reynolds, {\it et al.}: ``An integrated CMOS/memristor bio-processor for re-configurable neural signal processing," 2023 IEEE Biomedical Circuits and Systems Conference (BioCAS). IEEE (2023) (DOI: 10.1109/BioCAS58349.2023.10388703) 

\bibitem{website2}
Tsinghua University (2014) http://www.sigs.tsinghua.edu.cn/labs/vipl/thu-fvfdt.html.

\bibitem{journal paper17}
M. Chahardori, {\it et al.}: ``A 4-bit, 1.6 GS/s low power flash ADC, based on offset calibration and segmentation," IEEE Transactions on Circuits and Systems I: Regular Papers {\bf 60} (2023) 9 (DOI: 10.1109/TCSI.2013.2246206).

\bibitem{proceeding paper10}
H. J. Wu, {\it et al.}: ``A 1.2 V 4bit 25dB SNDR flash ADC with configurable output for GNSS receiver," 2018 14th IEEE International Conference on Solid-State and Integrated Circuit Technology (ICSICT). IEEE (2018) (DOI: 10.1109/ICSICT.2018.8564863)

\bibitem{journal paper18}
H. Y. Lee, {\it et al.}: ``500 MS/s 4-bit flash ADC with complementary architecture," Journal of Electromagnetic Engineering and Science {\bf 24} (2024) 1 (DOI: 10.26866/jees.2024.1.r.209).

\end{thebibliography}
\end{document}